\documentclass[twocolumn]{aastex63}

\usepackage{pgffor,upgreek,amsmath,nicefrac}
\usepackage{bm}
\graphicspath{{./}{figures/}}

\shorttitle{Non-thermal Emission from Cen~A jet}
\shortauthors{Sudoh et al.}

\renewcommand{\mu}{\upmu}

\makeatletter

\def\myd{\mathrm{d}}
\def\myd{d}
\def\dif{\@ifnextchar[{\@with}{\@without}}

\def\@with[#1]#2{
  \ensuremath{
    \mathchoice
    {\frac{\foreach \x in {#2}{\myd\x\,}}{\foreach \x in {#1}{\myd\x\,}}}%
    {{\foreach \x in {#2}{\myd\x\,}}/{\foreach \x in {#1}{\myd\x\,}}}%
    {{\foreach \x in {#2}{\myd\x\,}}/{\foreach \x in {#1}{\myd\x\,}}}%
    {{\foreach \x in {#2}{\myd\x\,}}/{\foreach \x in {#1}{\myd\x\,}}}
  }
}

\def\@without#1{
  \ensuremath{%
    \ifx\hfuzz#1\hfuzz
    \myd
    \else
    \foreach \x in {#1}{\myd\x\,}
    \fi
    }
}

\makeatother

\newcommand{\ergs}{\ensuremath{\mathrm{erg\,s^{{-1}}}}}
\newcommand{\ergscm}{\ensuremath{\mathrm{erg\,s^{-1}\,cm^{-2}}}}

\newcommand{\di}{\ensuremath{^{\textsc{d}}}}
\newcommand{\kn}{\ensuremath{^{\textsc{k}}}}
\newcommand{\vp}{\ensuremath{_{\textsc{v}}}}
\newcommand{\ic}{\ensuremath{_\textsc{ic}}}
\newcommand{\jet}{_{\rm jet}}
\newcommand{\keV}{\ensuremath{
    \mathchoice
    {_{\mathrm{ke}\textsc{v}}}%
    {_{\mathrm{ke}\textsc{v}}}%
    {\mathrm{ke}\textsc{v}}%
    {\mathrm{ke}\textsc{v}}
  }}
\newcommand{\TeV}{\ensuremath{\textsc{t}\mathrm{e}\textsc{v}}}
\newcommand{\Th}{\ensuremath{\textsc{t}{\scriptscriptstyle \mathrm{h}}}}
\newcommand{\NT}{\ensuremath{{\textsc{nt}}}}
\newcommand{\VHE}{\ensuremath{{\textsc{vhe}}}}
\newcommand{\B}{\ensuremath{
    \mathchoice
    {_\textsc{b}}%
    {_\textsc{b}}%
    {\textsc{b}}%
    {\textsc{b}}
}}

\begin{document}

\title{Physical Conditions and Particle Acceleration in the Kiloparsec Jet of Centaurus A}

\correspondingauthor{Takahiro Sudoh}
\email{sudoh@astron.s.u-tokyo.ac.jp}

\author[0000-0002-6884-1733]{Takahiro Sudoh}
\affil{Department of Astronomy, University of Tokyo, Hongo, Tokyo 113-0033, Japan}

\author[0000-0002-7576-7869]{Dmitry Khangulyan}
\affil{Department of Physics, Rikkyo University, Nishi-Ikebukuro 3-34-1, Toshima-ku, Tokyo 171-8501, Japan}

\author[0000-0002-7272-1136]{Yoshiyuki Inoue}
\affil{Interdisciplinary Theoretical \& Mathematical Science Program (iTHEMS), RIKEN, 2-1 Hirosawa, Saitama 351-0198, Japan}
\affil{Kavli Institute for the Physics and Mathematics of the Universe (WPI), UTIAS, The University of Tokyo, Kashiwa, Chiba 277-8583, Japan}

\begin{abstract}
\noindent The non-thermal emission from the kiloparsec-scale jet of Centaurus~A exhibits two notable features, bright diffuse emission and many compact knots, which have been intensively studied in X-ray and radio observations. H.E.S.S. recently reported that the very-high-energy gamma-ray emission from this object is extended along the jet direction beyond a kiloparsec from the core. Here, we combine these observations to constrain the physical conditions of the kpc-jet and study the origin of the non-thermal emission. We show that the diffuse jet is weakly magnetized ($\eta\B\sim10^{-2}$) and energetically dominated by thermal particles. We also show that knots are the sites of both amplified magnetic field and particle (re-)acceleration. To keep sufficient energy in thermal particles, the magnetic and non-thermal particle energy in the knot regions are tightly constrained. The most plausible condition is an energy equipartition between them, $\eta\B\sim\eta_e\sim0.1$. Such weak magnetic energy implies that particles in the knots are in the slow cooling regime. We suggest that the entire kpc-scale diffuse emission could be powered by particles that are accelerated at and escaped from knots.
\end{abstract}


\section{Introduction}

It is widely believed that jets from active galactic nuclei (AGN) are launched by electromagnetic mechanisms near supermassive black holes (SMBHs)~\citep[e.g.,][]{Blandford1977,Blandford1982,2007MNRAS.380...51K,McKinney2012}. As a result, jets are expected to be initially highly magnetized. The dissipation of the magnetic field converts the Poynting flux into the bulk kinetic energy, accelerating the jet to a relativistic speed. A fraction of the jet power is also transferred to particles, heating the jet material and accelerating particles to non-thermal energies. Particle acceleration can proceed effectively either in a magnetically-dominated (e.g., via magnetic reconnection) or kinetically-dominated (e.g., via formation of shocks) jet \citep[e.g.,][]{Sironi2015}. Therefore, to understand the production mechanism of non-thermal particles, the determination of the energy balance in the jets, especially their magnetization, is essential.

Observational studies of energy balance in AGN jets are mostly conducted for blazars, i.e., radio galaxies with their jets aligned toward Earth \citep[e.g.,][]{Tavecchio1998,2008MNRAS.385..283C,2010MNRAS.402..497G,Zhang2012,Inoue2016}. These studies typically find relatively weak magnetization. However, they are usually restricted to one-zone treatment aimed to explain observations at various phases. As blazars are highly variable \citep[e.g.,][]{Ackermann2016}, it is unclear whether the emission of each phase correctly probes conditions in the large-scale jet. Most blazars are located at cosmological distances, which makes any study beyond one-zone treatment difficult.

The radio galaxy Centaurus A (Cen A) enables invaluable insights on this problem thanks to its unequaled proximity~\citep[3.8~Mpc;][]{Harris10}. Broadband emission from this object has been resolved over a wide range of spatial scales from the core ($\lesssim 10^{-2}$~pc) to the giant lobes ($\gtrsim 10^5$~pc) \citep[e.g.,][]{Kraft02,Hardcastle06,Kataoka06,Goodger10}. Recently, the H.E.S.S. collaboration has reported evidence of very-high-energy (VHE) gamma rays from the kpc-scale jet in Cen A~\citep{nature}. A combination of new gamma-ray data with previous multi-wavelength data brings new information on jet properties on kpc distances from SMBH.

Here, we study the origin of non-thermal emission and physical conditions of the kpc-jet in Cen A. Our approach is model-independent, meaning that we rely on observational data only. In Sec.~\ref{sec:obs}, we summarize observational properties. In Sec.~\ref{sec:xray}, we constrain physical conditions in the jet required from X-ray observations. In Sec.~\ref{sec:VHE}, we further constrain the parameter space with the VHE data. In Sec.~\ref{sec:discuss}, we summarize our findings. 

\section{Observational Properties}
\label{sec:obs}
The SMBH at the core of Cen A has a mass of $5.5\times10^7~{\rm M}_\odot$~measured by stellar kinematics \citep{Cappellari09}, with corresponding Eddington luminosity of $7\times 10^{45}\,\ergs$. It provides an ultimate energy source to the jet, which has an estimated power of $\sim 10^{43}\,\ergs$\citep{Wykes13}, an apparent velocity of $\simeq0.5c$ \citep{hardcastle2003} on a hundred-parsec scale, and an opening angle of $10^\circ-15^\circ$ \citep[e.g.,][]{Horiuchi06}. 

On kpc scales, the jet produces diffuse synchrotron emission. \cite{Kataoka06} utilized {\it Chandra} data and obtained the X-ray flux along the jet from the core up to about 240${''}$ (4~kpc). The observed 0.5$-$5 keV luminosity of the diffuse unresolved kpc-scale jet is $L_{\keV}\di \simeq8\times10^{38}\,\ergs$, where the superscript $\textsc{d}$ stands for the diffuse component. We define this energy range as the keV~band. The spectral index of this component, $\alpha
=-\dif[\ln\nu]{\ln F_\nu}$, is consistent with $\alpha\simeq1$. 

The jet contains individual knots resolved in X-ray and radio observations~\citep{Kraft02,Goodger10}. The number of X-ray knots identified in \cite{Kataoka06} is about 30. While $\sim2-5$ of them could be low-mass X-ray binaries unrelated to the jet emission~\citep{Goodger10}, the majority are produced by the jet material~\citep{1979ApL....20...15B,1983ApJ...266...73S,hardcastle2003,2007ApJ...669L..13M,Bednarek15,2017A&A...604A..57V,Torres-Alba19}. The typical keV-band luminosity of each knot is $L_{\keV}\kn\simeq10^{37}\,\ergs$, where the superscript $\textsc{k}$ stands for knots. The X-ray spectral indices are consistent with $\alpha\simeq0.5-1$~\citep{Goodger10,Tanada19}. For some knots, the spectral indices in radio band (4.8$-$8.4~GHz) are also measured, in the range of $\alpha\simeq0.5-2$, although uncertainties are large~\citep{Goodger10}.

The sizes of knots are constrained only for some of the brightest ones, typically $\simeq2-10$~pc~\citep{Tingay09,Goodger10,Tanada19}. The magnetic fields in the knots are also largely unconstrained. For two bright knots, BX2 and AX1C, {\it Chandra} observations suggest upper limits of $B\lesssim80~\mu$G due to the absence of spectral steepening expected for synchrotron cooling~\citep{Snios19}.

The production of synchrotron emission in the energy of $\epsilon=1\epsilon_1$~keV implies the presence of electrons with energies of $E_e\simeq$10$\epsilon_1^{\nicefrac12}B_{100}^{\nicefrac{-1}2}$~TeV, where $B=100B_{100}$~$\mu$G is the magnetic field strength. The same population of electrons produces gamma rays by inverse Compton (IC) scattering. If the scattering proceeds in the Thomson regime, the characteristic gamma-ray energy is
\begin{equation}
    \epsilon\ic  = 300~\left({\frac{\hbar\omega_0}{6\times 10^{-4}\rm\,eV}}\right)\left({\frac{E_e}{10\rm\,TeV}}\right)^{2}~\rm GeV,
\end{equation}
where $\hbar\omega_0$ is the energy of target photons and the Thomson limit is valid for $\hbar\omega_0\ll 0.1 \left(E_e/10{\rm \,TeV}\right)^{-1}\rm\,eV$.
The recent H.E.S.S. analysis has confirmed that VHE emission is produced in the kpc-jet~\citep{nature}. The flux is approximately $2\times 10^{-10}$~GeV~cm$^{-2}$~s$^{-1}$ at $\epsilon\ic \simeq300$~GeV and the spectrum is fit by a power-law with $\alpha\simeq1.5$ up to $\sim10$~TeV~\citep{HESS18}. Thus, the luminosity in the VHE band, which we define as 0.3$-$3~TeV, is $L_{\VHE}\simeq 7\times 10^{38}\,\ergs$.

\begin{table}
\begin{center}
     \caption{Luminosity of the host galaxy NGC 5128}
\begin{tabular}{ccc} \hline
     Band & Wavelength & Luminosity \\ 
      & [$\mu$m]& [erg~s$^{-1}$]  \\ \hline
     optical & 0.6 &$\sim$9$\times$10$^{43}$  \\ 
     near-IR & 2 & $\sim$6$\times$10$^{43}$  \\
     mid-IR & 25 & $\sim$0.6$\times$10$^{43}$  \\
     far-IR & 100 & $\sim$2$\times$10$^{43}$ \\ \hline
\end{tabular}
     \label{table:luminosity}
\end{center}
\end{table}

The target photon fields may be produced by the jet itself, objects in it, and external sources. The host galaxy provides the brightest external soft photons in optical and infrared. Table~\ref{table:luminosity} shows their characteristics taken from NASA/IPAC Extragalactic Database. Note that soft photons with shorter wavelengths are affected by Klein--Nishina suppression. For example, for an electron spectrum of $\dif[E_e]{N_e}\propto E_e^{-3}$, the contributions from optical and near-IR are smaller than that of far-IR above $\epsilon\ic\simeq100$~GeV, despite the higher luminosities. The nucleus of Cen A could also provide target photons with a bolometric luminosity of $\sim10^{43}\,\ergs$ \citep{Chiaberge01,Beckmann11}.

Although gamma rays can be produced also by hadronic processes, their contributions are likely small~(Appendix~\ref{app:hadron}). Throughout, we assume that VHE gamma rays are predominately generated by leptons.\\ \\

\section{Constraints from X-ray data}
\label{sec:xray}
\subsection{Jet Energy Balance}
The average physical conditions in the jet are determined by its basic properties. We assume that the kpc jet is cylindrical with a radius $R$ and height $Z=3Z_3$~kpc, starting from a distance of 1~kpc from the core. We use an opening angle of $\theta=0.2\theta_{0.2}$~rad, which results in $R = 200\theta_{0.2}$~pc. This might appear an overestimate for the radius at 1~kpc~\cite[e.g.,][]{Wykes19}. However, for our cylindrical approximation, it would be appropriate as the mean radius of the kpc jet. We assume a total jet power of $P\jet= 10^{43}P_{43}\,\ergs$. The energy flux in the jet is
\begin{equation}\label{eq:jet_flux}
    \frac{P\jet}{\pi R^2} \simeq 8\,P_{43}\theta_{0.2}^{-2}~{\ergscm}\,.
\end{equation}
The jet bulk speed, $\beta=0.5\beta_{0.5}$, defines the energy density
\begin{equation}\label{eq:jet_w}
    w\jet=\frac{P\jet}{\pi R^2\beta c} \simeq350\,P_{43}\theta_{0.2}^{-2}\beta_{0.5}^{-1}~{\rm eV~cm^{-3}}\,,
\end{equation}
which is distributed to thermal gas, magnetic field, and non-thermal protons and electrons, such that
\begin{equation}\label{eq:jet_composition}
    w_{\Th}+w\B +w_{e,\NT}+w_{p,\NT}=w\jet\,.
\end{equation}
We define the corresponding fractions, $\eta_i=w_i/w\jet$:
\begin{equation}\label{eq:jet_fractions}
    \eta_{\Th}+\eta\B +\eta_{e,\NT}+\eta_{p,\NT}=1\,.
\end{equation}
The magnetic energy density, 
\begin{equation}
  w\B  = \frac{B^2}{8\pi} = \eta\B w\jet\,,
\end{equation}
converts to the strength of the magnetic field:
\begin{equation}
  \label{eq:mag}
  B = \sqrt{\frac{8\pi \eta\B  P_j}{\pi R^2 \beta c}} = 120\sqrt{\frac{\eta\B P_{43}}{\beta_{0.5}}}\theta_{0.2}^{-1}\ {\rm \mu G}\,.  
\end{equation}

The energy distribution for non-thermal particles is often approximated with a broken power law:
\begin{equation}
  \dif[E,V]{n_{Y}}=\left\{
    \begin{matrix}
      A_Y\left(E/E_{Y,\rm br}\right)^{-p_{Y, H}}& E\geq E_{Y,\rm br}\,,\\
      A_Y\left(E/E_{Y,\rm br}\right)^{-p_{Y, L}}& E_{Y,\rm min}<E\leq E_{Y,\rm br}\,.\\
    \end{matrix}
    \right.
  \end{equation}
Here $Y$ denotes the particle type, $p_{Y,L}$ and $p_{Y,H}$ define power-law slopes,  $E_{Y,\rm br}$ is the break energy, $A_Y$ is the normalization, and $E_{Y,\rm min}$ is the minimum energy of the non-thermal distribution. The energy density in non-thermal particles is
\begin{equation}
    w_{Y,\NT} = \int\limits_{E_{Y,\rm min}}^{\infty} \dif[E,V]{n_{Y}}E\dif{E}\,.\\
\end{equation}

For electrons, radio-emitting particles typically have low energies ($E_e<E_{e,\rm br}$), while X-ray emitting particles have high energies ($E_e>E_{e,\rm br}$). Therefore, the energy content in non-thermal electrons that produce emission in the keV-band is
\begin{equation}
    w_{e,\keV} =  A_e\int\limits_{\sqrt{0.5}E_{e, \keV }}^{\sqrt{5}E_{e, \keV }} \left(\frac{E}{E_{e,\rm br}}\right)^{-p_{e,H}}E\dif{E}\,,
\end{equation}
where $E_{e, \keV }$ is the energy of electrons which are responsible for the production of $1~\rm\,keV$ synchrotron photons: $E_{e, \keV }\simeq10 \left(B/100~\mu{\rm G}\right)^{\nicefrac{-1}2}\rm\,TeV$. It is useful to define another parameter,
\begin{equation}
\label{eq:chi}
    \chi\keV  = \frac{w_{e,\keV }}{w_{e,\NT}} = \frac{\eta_{e,\keV }}{\eta_{e,\NT}},
\end{equation}
which is determined by the electron spectrum.

\subsection{X-ray Emission from the Jet}

X-ray observations of Cen A revealed both diffuse unresolved emission and many compact knots in the kpc jet. The diffuse jet luminosity in the keV band is $L_{\keV}\di\simeq 8\times10^{38}~{\ergs}$ \citep{Kataoka06}, and the volume is $V\di\simeq\pi R^2 Z\simeq10^{64}\theta_{0.2}^2Z_3\rm \,cm^3$. The luminosity density, $j=L/V$, of the diffuse jet is
\begin{equation}\label{eq:diffuse_emis}
  j_{\keV}\di = 7\times10^{-26}\theta_{0.2}^{-2}Z_3^{-1}\rm\, \ergs cm^{-3}
\end{equation}

The compact knots have a typical keV-band luminosity of $L_{\keV}\kn=10^{37}L_{37}\,\ergs$. We adopt a characteristic knot size of $r=5r_5$~pc and assume that knots are spherical. Then, the typical volume is $V\kn=2\times10^{58}r_5^3\rm\,cm^3$. The luminosity density of the knot radiation is
\begin{equation}\label{eq:knot_emis}
  j_{\keV}\kn=6\times10^{-22}r_5^{-3}L_{37}\rm\, \ergs cm^{-3}\,.
\end{equation}

These luminosities are determined by the energy content in keV-emitting electrons and their synchrotron cooling time,
\begin{equation}
    t_{\rm syn} 
    \simeq 2\times10^2 \left(\frac{\hbar \omega}{1\,\rm keV }\right)^{\nicefrac{-1}2}\left(\frac{w\B }{100\,\rm eV\, cm^{-3}}\right)^{\nicefrac{-3}4}\rm\,yr\,,\\
\end{equation}
where $\hbar\omega$ is the synchrotron emission energy. By equating $j\keV $ with $w_{e,\keV }/t_{\rm syn}$, we obtain the following constraints on the production sites of synchrotron emission:
\begin{equation}
\label{eq:diffuse}
    (\eta_{\B}\di )^{\nicefrac34}\eta_{e,\keV }\di  \simeq 3\times10^{-7}\theta_{0.2}^{\nicefrac32}Z_3^{-1}\beta_{0.5}^{\nicefrac74}P_{43}^{\nicefrac{-7}4}\,
\end{equation}
and
\begin{equation}
\label{eq:knot}
    (\eta_{\B}\kn)^{\nicefrac34}\eta_{e,\keV }\kn r_5^3 \simeq 3\times10^{-3}L_{37}\theta_{0.2}^{\nicefrac72}\beta_{0.5}^{\nicefrac74}P_{43}^{\nicefrac{-7}4}\,.
\end{equation}

\begin{figure}[b]
    \centering
    \includegraphics[width=\columnwidth]{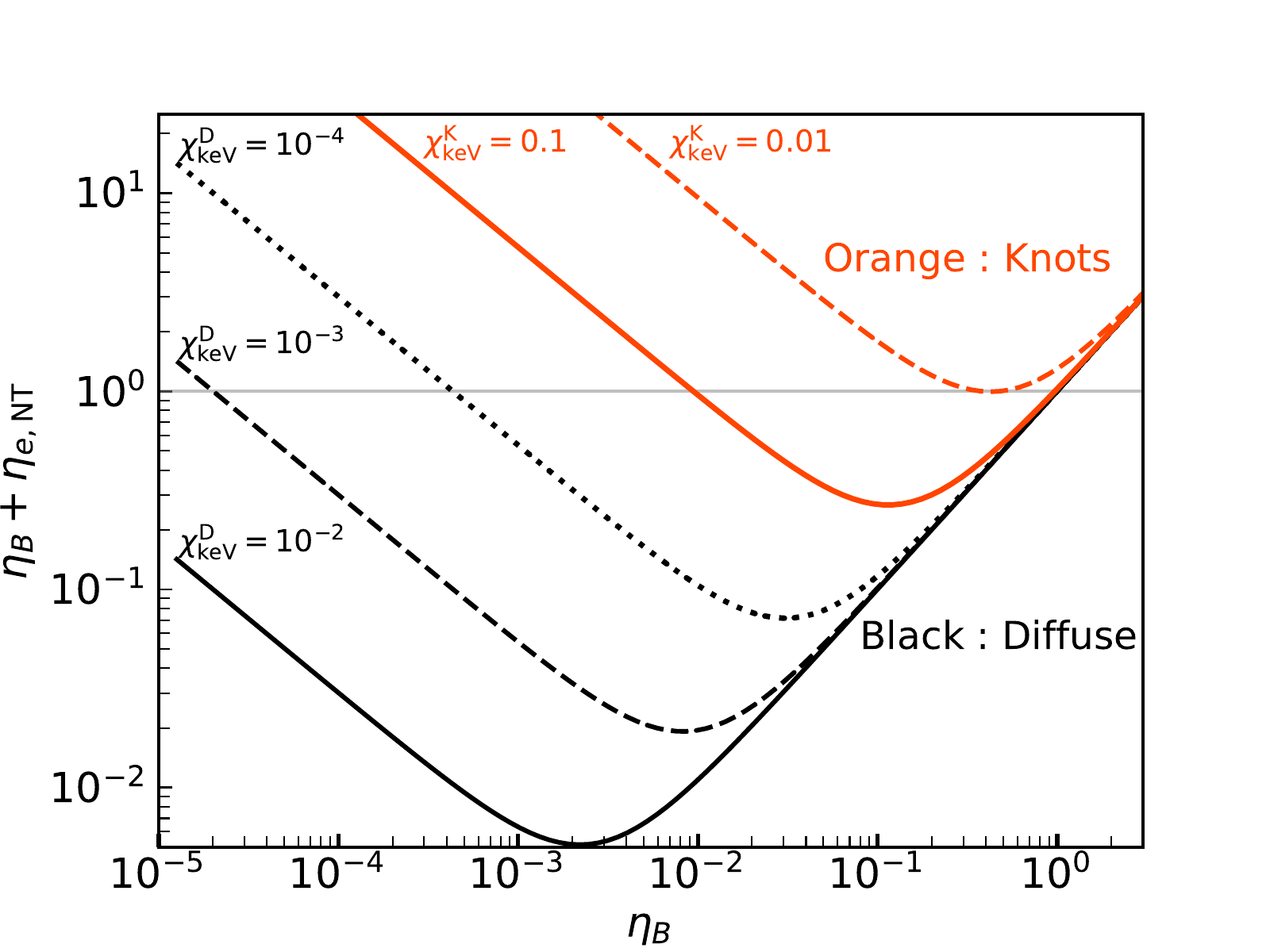}
    \caption{The sum of non-thermal electron and magnetic energy as a function of $\eta\B $, under the constraints from X-ray observation: Eq.~(\ref{eq_diff}) for the diffuse jet and Eq.~(\ref{eq_knot}) for the knots. This sum cannot (significantly) exceed one, which constrains these parameters.}
    \label{fig:eta_sum}
\end{figure}

To relate $\eta_{e,\keV }$ to $\eta_{e, \NT}$, we need the electron spectrum. In general, the ratio of these two, $\chi\keV $, is larger for a harder spectrum. For the diffuse jet, \citet{Hardcastle11} derived $p_{e,L}\simeq2.06$ for radio-emitting electrons and a very steep spectrum for X-ray emitting particles, $p_{e,H}\simeq3.88$ and $E_{e,\rm br}\sim 10^{-1.5}E_{e,\keV }$, based on multi-wavelength data~\citep{Hardcastle06}. If we assume $E_{e,\rm min}\sim 10^{-3}E_{e,\rm br}$, these values convert to $\chi_{\keV}\di \sim10^{-4}$. We note that this spectrum is obtained from about 2$-$4 arcmin (2$-$4~kpc) from the core. Closer to the core, the X-ray spectrum may be harder. Indeed, the X-ray spectrum derived in \cite{Kataoka06} for the 1$-$2~kpc jet yields $p_{e,H}\simeq3.0-3.4$, resulting in $\chi_{\keV}\di \sim 10^{-3}$. Therefore, the value of $\chi_{\keV}\di$ averaged over the kpc-jet is likely larger than $10^{-4}$. Utilizing this, we rewrite Eq.~(\ref{eq:diffuse}) as
\begin{equation}
\label{eq_diff}
    (\eta_{\B}\di )^{\nicefrac34}\eta_{e,\NT}\di  \simeq 3\times10^{-3}\theta_{0.2}^{\nicefrac32}Z_3^{-1}\beta_{0.5}^{\nicefrac74}P_{43}^{\nicefrac{-7}4}\chi_{-4}^{-1},
\end{equation}
where $\chi_{-4}=\chi_{\keV}\di /10^{-4}$. We constrain the jet parameters by limiting the sum of non-thermal electron and magnetic energy, $\eta_{\B}\di+\eta_{e,\NT}\di<1$. Figure~\ref{fig:eta_sum} shows this sum under the above constraint (Eq.~\ref{eq_diff}) as a function of $\eta_{\B}\di$. This shows that a wide range of $\eta_{\B}\di$ and $\eta_{e,\NT}\di$ is allowed from the X-ray data.

The spectral properties of knots are less tightly constrained from observations and may differ from one to another. However, observed X-ray fluxes can place useful constraints. To illustrate this, we write Eq.~(\ref{eq:knot}) as
\begin{equation}
\label{eq_knot}
    (\eta_{\B}\kn )^{\nicefrac34}\eta_{e,\NT}\kn  \simeq 3\times10^{-2}L_{37}\theta_{0.2}^{\nicefrac72}\beta_{0.5}^{\nicefrac74}P_{43}^{\nicefrac{-7}4}r_5^{-3}\chi_{-1}^{-1}\,,
\end{equation}
where $\chi_{-1}=\chi_{\keV}\kn /10^{-1}$. Figure~\ref{fig:eta_sum} also shows $\eta_{\B}\kn+\eta_{e,\NT}\kn$ under the above constraint. Because this sum should not significantly exceed one, we obtain $\chi_{\keV}\kn \gtrsim0.01$. This places tight constraints on the spectrum. If we use $p_{e,L}\simeq2.3$ as an example, then $E_{e,\keV}$ should be nearly equal to (or smaller than) $E_{e,\rm br}$ to satisfy this constraint. If we instead use $p_{e,L}\simeq2.06$ as derived for the diffuse jet, $E_{e,\keV}$ still need be relatively close to the break: $E_{e,\keV} \lesssim 10~E_{e,\rm br}$. In the latter case, $\chi_{\keV}\kn $ can be as high as 0.1.

\subsection{Conditions in the Jet}
X-ray observations allow three distinct scenarios for the physical conditions in the diffuse jet.
\begin{itemize}
\item[(i)] Strongly magnetized, $\eta_{\B}\di \sim1$
\item[(ii)] Non-thermal electron dominant, $\eta_{e,\NT}\di \sim1$
\item[(iii)] Thermal plasma dominant, $\eta_{\Th}\di \sim1$
\end{itemize}

{\it Case (i):} If the entire jet is strongly magnetized, the knot would also have a high magnetization, $\eta_{\B}\kn \sim 1$. The large difference in luminosity densities would be due to higher electron energy densities in the knots, implying efficient particle (re-)\-acceleration taking place there.

{\it Case (ii):} If the jet total energy is mostly carried by non-thermal electrons, the diffuse jet should be very weakly magnetized: $\eta_{\B}\di \sim 4\times10^{-4}~(\chi_{-4})^{\nicefrac{-4}3}$~(Eq.~\ref{eq_diff}). In this case, amplification of the magnetic field would be needed to explain the compact knots. 

{\it Case (iii):} If the bulk of the jet energy is carried by thermal particles, the jet magnetization would be relatively weak because $\eta_{\B}\di  + \eta_{e,\NT}\di \ll 1$. In the knot region, we would expect larger values of $\eta_{\B}\kn $ and $\eta_{e,\NT}\kn $, but the thermal particles should still have most of the energy there. A plausible realization of such a scenario is equipartition between the magnetic field and the non-thermal electrons, $\eta_{\B}\kn \simeq\eta_{e,\NT}\kn$, which minimizes the energy requirement for these two components.

Next, we show that only case (iii) is allowed by the VHE data.

\section{Further Constraints from VHE data}
\label{sec:VHE}
\subsection{VHE Emission from the Jet}

Here, we combine X-ray and gamma-ray data to further constrain the jet properties. To do so, we note that X- and gamma-ray instruments have different angular resolution. In particular, the VHE flux includes contributions both from the jet and counter-jet. The emission from the counter-jet is Doppler de-boosted by a relative factor of $\left({\cal D}_{\rm cj}/{\cal D}\right)^2$ \citep[see, e.g.,][]{2018MNRAS.481.1455K}. Here ${\cal D}_{\rm cj}=1/(\Gamma(1-\beta\cos(\uppi-\theta_{\rm obs})))$ and ${\cal D}=1/(\Gamma(1-\beta\cos\theta_{\rm obs}))$ are the Doppler factors for the counter-jet and jet, respectively. The bulk Lorentz factor, $\Gamma=1/\sqrt{1-\beta^2}$, is assumed to be common and $\theta_{\rm obs}$ is the angle between the light-of-sight and the jet velocity. The Doppler de-boosting is significant, ranging between $0.07$ and $0.3$ for feasible viewing angles of $20^\circ-50^\circ$~\citep[][see also Appendix.~\ref{app:beaming}]{Tingay1998,hardcastle2003}. 

However, the counter-jet VHE emission can be still important since IC may proceed there at a more favorable scattering angle. Below, we formally ignore the contribution from the counter-jet and use an isotropic approximation for the target photons. If they are indeed isotropic, we overestimate the VHE emission from the jet by the factor of $1+\left({\cal D}_{\rm cj}/{\cal D}\right)^2$, i.e., our estimate will be accurate within a factor of 1.3.

The photon field may instead originate in the core region. Then, for a power-law distribution of electrons with index $p$ emitting in the Thomson regime, the flux is reduced compared to the isotropic case by a factor of \(A(\theta)=\left(1-\cos\theta\right)^{\nicefrac{p+1}2}\) \citep[see, e.g.,][here for simplicity we adopted the scattering at the angle of \(90^\circ\) as an estimate for the isotropic case]{2018MNRAS.481.1455K}. In such a case, the emission from the jet is strongly suppressed and the VHE emission is produced in the counter-jet. Then we underestimate the total VHE emission from the jet by a factor of $A\left(\theta_{\rm obs}\right) + A\left(\uppi-\theta_{\rm obs}\right)\left({\cal D}_{\rm cj}/{\cal D}\right)^2\,.$ For a cooled electron spectrum, $p=3$, this yields $0.3$ and $0.9$ for the viewing angle of $20^\circ$ and $50^\circ$, respectively. 

The above estimates indicate that our simple consideration is accurate within a factor of $3$ independently of the angular distribution of the photons, leaving us with the target energy density, $w_{\rm ph}$, as the only parameter for our estimates.

We assume that the VHE emission is produced by IC scattering in the Thomson regime, which is valid if the target photon is dominantly provided by far-IR radiation. The observed keV and VHE luminosity relates as $L\keV/L_{\VHE}=w\B/w_{\rm ph}$. Therefore, the following conditions are required, depending on the production site of VHE emission: 
\begin{equation}
  \begin{split}
    w_{\rm ph} &= \frac{L_{\VHE}}{L_{\keV}\di}w_{\B}\di \simeq w_{\B}\di\rm \ (diffuse\ jet), \\
    w_{\rm ph} &= \frac{L_{\VHE}}{NL_{\keV}\kn}w_{\B}\kn \simeq 2w_{\B}\kn\rm \ (knots),
  \end{split}
\end{equation}
where $N(\simeq30)$ is the number of knots. 

Since knots have much higher synchrotron luminosity density than the diffuse jet, we would naturally expect $w_{\B}\kn \geq w_{\B}\di $. Then, if VHE emission is dominated by knots, they should have locally enhanced photon fields. To be relevant, the knot additional photon field should have an energy density comparable to that of the magnetic field. The luminosity should be
\begin{equation}
  \begin{split}
    L_{\rm ph}^{\rm add}&= 4\pi r^2 cw_{\B}\kn \,, \\
    &\simeq 5\times10^{40} \eta_{\B}\kn r_5^{2}P_{43}\theta_{0.2}^{-2}\beta_{0.5}^{-1}\rm\, \ergs\,.
  \end{split}
\end{equation}
Since the knot magnetic field is $\eta_{\B}\kn \gtrsim10^{-2}$ from X-ray observation (Figure~\ref{fig:eta_sum}), a luminous photon source brighter than $5\times 10^{38}~\ergs$ would be needed. This is much brighter than the X-ray luminosity of each knot, which indicates that synchrotron self-Compton cannot be sufficient. In principle, luminous stars could provide this photon field, but in that case, the production of VHE should proceed in the Klein-Nishina regime, significantly decreasing the efficiency of the IC process. Therefore, we regard this possibility as unlikely and consider the diffuse jet as the origin of the VHE emission.

If the target photons are generated inside the jet, the required photon luminosity is 
\begin{equation}\label{eq:internal_photons}
   \pi RZc w_{\rm ph} = 3\times 10^{44} \eta_{\B}\di\frac{P_{43}Z_3}{\theta_{0.2}^2}\ {\ergs}\,.
\end{equation}
This scenario necessitates a relatively small jet magnetization, $\eta_{\B}\di\lesssim0.03$, because otherwise the required luminosity would exceed the total jet power. The required luminosity can be decreased by a factor of $R/Z\sim0.2$, if we assume that the target photons are strongly beamed along the jet axis \citep[e.g., if emitted by the highly relativistic inner jet, see][]{Bednarek18}, but this still requires $\eta_{\B}\di$ well below $0.1$.

\begin{figure}
    \centering
    \includegraphics[width=\columnwidth]{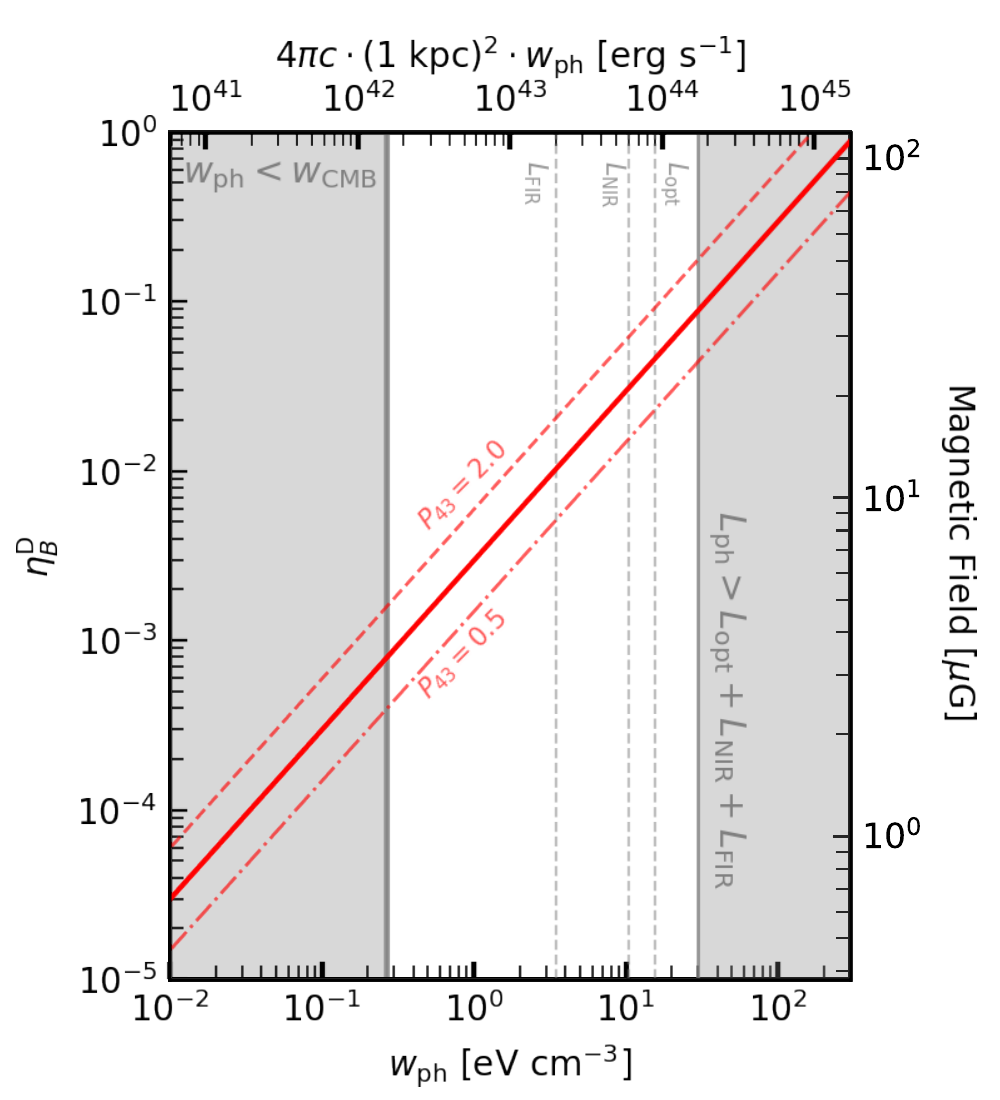}
    \caption{Jet magnetization that is required for explaining VHE emission, as a function of soft photon energy density. The solid line adopts $P_{\rm 43}=1$, while other two lines show the cases when $P_{\rm 43}$ is changed to 2 (dashed) and 0.5 (dot-dashed). Other model parameters are fixed to the value as adopted in the main text.}
    \label{fig:diffuse}
\end{figure}

If the soft photons are supplied from external sources, the luminosity should be 
\begin{equation}
    w_{\rm ph}4\pi d^2c = 2\times 10^{45}\eta_{\B}\di\frac{P_{43}d_1^2}{\theta_{0.2}^2\beta_{0.5}}\ {\ergs}
\end{equation}
where $d=1d_1$ kpc is the distance from the emitting region to the source of soft photons. We note that the use of a single parameter $d$ is a simplification, because external photon sources (starlight and dust) are spatially extended. A more realistic calculation by \cite{Stawatz06} finds an energy density of $\sim10^{-11}$~erg~cm$^{-3}$ from stars in the kpc-jet (see their Figure 2), consistent with $d\sim1$~kpc in our estimate. This scenario necessitates a relatively small jet magnetization, $\eta_{\B}\di\lesssim0.03$, because otherwise the luminosity of the host galaxy would not be sufficient.

Both internal and external sources of target photons require $\eta_{\B}\di <0.03$. This excludes the possibility of a strongly magnetized jet (case i). Figure~\ref{fig:diffuse} shows the jet magnetization parameter in the diffuse jet as a function of the photon energy density. It also shows that the jet magnetization should satisfy $\eta_{\B}\di \gtrsim10^{-3}$, because otherwise the emission produced on the cosmic microwave background radiation would be brighter than observed. This excludes the case of a very weakly magnetized jet (case ii). Adopting a luminosity of $\simeq10^{44}~\ergs$, we find $\eta_{\B}\di \simeq 5\%$ and $\eta_{e,\rm NT}\di \simeq 3\%$. We conclude that the thermal plasma dominates the jet energetics (case iii).

Thus far, we have shown that thermal particles carry most of the energy in the diffuse jet. It may not seem straightforward to relate this with the energetics in knots, because they locally have very different conditions. However, considering that knots are likely produced by the same material that composes the diffuse jet, in which non-thermal electrons and magnetic field make only $\simeq8\%$ of the energy in total, thermal plasma is likely dominant also in the knot regions. In principle, we cannot rule out a possibility that magnetic fields are amplified to $\eta_{\B}\kn\simeq\eta_{\rm Th}\kn$, but it would require very efficient conversion of thermal energy to the magnetic field. A more feasible scenario is to keep most of the energy in thermal particles in the knot region by minimizing $\eta_{\B}\kn+\eta_{e,\NT}\kn$ (i.e., equipartition), with $\eta_{\B}\kn$ and $\eta_{e,\NT}\kn$ amplified by a factor of $\mathcal{O}(1)$ compared to the diffuse jet. This argument favors $\eta_{\B}\kn \simeq \eta_{e,\NT}\kn \sim0.1$ (see Fig.~\ref{fig:eta_sum}). This results in a relatively weak magnetic field, $B\kn\sim40~\mu{\rm G}$, consistent with the observed upper limits for two bright knots~\cite[$B\kn<80~\mu{\rm G}$;][]{Snios19}.

\subsection{Implications for Particle Acceleration}
The dominance of thermal particles implies a relatively small magnetic field even in the knot region. Particles produced in knots can travel a distance of
\begin{equation}
  \begin{split}
    r_{\rm syn} &= c\beta t_{\rm syn}\kn\,,\\
    &\simeq60 \left(\frac{\hbar \omega}{1\,\rm keV }\right)^{\nicefrac{-1}2}\left(\frac{\eta_{\B}\kn}{0.1}\right)^{\nicefrac{-3}4}P_{43}^{\nicefrac{-3}4}\theta_{0.2}^{\nicefrac32}\beta_{0.5}^{\nicefrac74}\rm\,pc\,.
  \end{split}
\end{equation}
before losing energy to the synchrotron cooling. This is significantly larger than typical knot size, which indicates that they escape from the knots and contribute to the diffuse emission. The electron power supplied by escaping particles from each knot is $L_{\keV}\kn t_{\rm syn}\kn/t_{\rm adv}\kn$, where $t_{\rm adv}\kn =r/\beta c$ is the advection time in the knot. They would cool down in the diffuse jet, radiating with an X-ray luminosity of
%
\begin{equation}
 \begin{split}
  L\keV &\sim\frac{\xi}{2}NL_{\keV}\kn \frac{t_{\rm syn}\kn }{t_{\rm adv}\kn } \\
  &\sim90L_{\keV}\kn\left(\frac{\xi}{0.5}\right)\left(\frac{N}{30}\right)P_{43}^{\nicefrac{-3}4}r_5^{-1}\beta_{0.5}^{\nicefrac74}\theta_{0.2}^{\nicefrac32} \\
  &\sim9\times10^{38}~\ergs\,,
  \end{split}
\end{equation}
where $1/2$ roughly accounts for the synchrotron and IC cooling. The parameter $\xi$ takes the difference in magnetic fields between diffuse jet and knots: since $E_{e, \keV}\propto(\eta_B)^{-1/4}$, the energy in keV-emitting electrons differ by a factor of $\xi\sim(\eta_B\kn/\eta_B\di)^{(2-p_H)/4}\sim0.5$. This estimate is remarkably close to the diffuse luminosity, suggesting that the particles accelerated in the compact knots may play an essential role in producing the jet diffuse emission. 

\section{Conclusion}
\label{sec:discuss}

In this work, we study the origin of non-thermal emission and physical conditions in the kpc-jet of Cen A. By combining X-ray and VHE data, we determine the jet magnetization to be $\eta_{\B}\di \sim10^{-2}$ in the kpc-jet. This result is consistent with a recent study on FR~II radio galaxies~\citep{Sikora20}. In knot regions, the energy densities in the magnetic field and non-thermal electrons should be amplified to an equipartition value, $\eta_{\B}\kn \simeq\eta_{e,\NT}\kn \sim0.1$. Such a weak magnetic field implies that most particles leave knots uncooled. We find that it remains viable that entire jet X-ray and VHE emission is produced by particles that are accelerated at and escaped from knots. More detailed modeling is needed to test this scenario.

\section*{Acknowledgements}
We thank Felix Aharonian, Maxim Barkov, Valent\'i Bosch-Ramon, Jun Kataoka, Frank Rieger, and Marek Sikora for useful comments. We also thank the anonymous referee for helpful comments. This research made use of the NASA/IPAC Extragalactic Database (NED), which is operated by the Jet Propulsion Laboratory, California Institute of Technology, under contract with the National Aeronautics and Space Administration. T.S. is supported by Research Fellowship of Japan Society for the Promotion of Science (JSPS), and also supported by JSPS KAKENHI Grant Number JP 18J20943. Y.I. is supported by JSPS KAKENHI Grant Number JP16K13813, JP18H05458, JP19K14772, program of Leading Initiative for Excellent Young Researchers, MEXT, Japan, and RIKEN iTHEMS Program. D.K. is supported by JSPS KAKENHI Grant Numbers JP18H03722, JP24105007, and JP16H02170.

\appendix
\vspace{-0.3cm}
\section{Relativistic Beaming}
\label{app:beaming}
Since the jet bulk velocity is only mildly-relativistic, we do not consider relativistic effects in the main text. Here we discuss their possible impact.

Relativistic effects depend on the jet bulk velocity, $\beta$, and the viewing angle, $\theta_{\rm obs}$, through the Lorentz and Doppler factors, $\Gamma=1/\sqrt{1-\beta^2}$ and ${\cal D}=1/(\Gamma(1-\beta\cos\theta_{\rm obs}))$. The viewing angle is observationally uncertain, probably being in the range of $\theta_{\rm obs}\simeq20^\circ-50^\circ$~\citep{Tingay1998,hardcastle2003}. This also induces some uncertainties in the jet velocity, because it is obtained from the viewing angle and the apparent velocity $\beta_{\rm app}$ as 
\begin{equation}
  \beta = \frac{\beta_{\rm app}}{\beta_{\rm app}\cos\theta_{\rm obs}+ \sin\theta_{\rm obs}}\,. 
\end{equation}
If we fix the jet apparent velocity to $\beta_{\rm app}\simeq0.5c$ \citep{hardcastle2003} and use $20^\circ < \theta_{\rm obs} < 50^\circ$, then the Lorentz factor $\Gamma$ is at most 1.3, and ${\cal D}$ is smaller than 2. While these suggest that relativistic effects are not critical, they may deserve a more careful check because they could strongly depend on $\Gamma$ and ${\cal D}$.

To correctly account for relativistic effects, we have three points to be modified. First, the jet energy density in the jet comoving frame should be $P\jet/(\pi R^2\Gamma^2\beta c)$, i.e., the definition in Eq.~(\ref{eq:jet_w}) needs to be modified by a factor of $\Gamma^{-2}$. Other energy parameters $w_i$ and $\eta_i$ are defined in the jet comoving frame in the main text. Second, the X-ray luminosity density $j\di_{\keV}$ is calculated in the observer frame, since it is estimated from the observed X-ray flux and volume. To transform $j\di_{\keV}$ to the comoving frame, we use a well-known fact that ${j_\nu}/{\nu^2}$ is a Lorentz invariant, where we adopt the standard notation of $j_\nu={dE}/{(dVdtd\Omega d\nu)}$ \citep[e.g.,][]{1979book}. Therefore, $j_\nu$ transforms as ${\cal D}^2$. Since we consider an isotropic emitter, $j\keV$ is related to $j_\nu$ as $ j_\nu d\nu = {j\keV}/{4\pi}$. This indicates that the X-ray luminosity density in the jet comoving frame is ${\cal D}^{-3}j\di_{\keV}$. Third, since the synchrotron cooling time ($t_{\rm syn}$) is proportional to $\nu^{-1/2}B^{-3/2}$, in the comoving frame it is longer by a factor of $D^{1/2}\Gamma^{3/2}$. Combining these three points, the energy density of electrons ($w\di_{e,\keV}$) should be proportional to ${\cal D}^{-5/2}\Gamma^{3/2}$, and, the parameter $\eta_{e,\NT}\di$ should change as
\begin{equation}\label{eq:rel_factor}
\eta_{e,\NT}\di \propto {\cal D}^{-5/2}\Gamma^{7/2}
\end{equation}
If we fix $\beta_{\rm app}\simeq0.5c$~\citep{hardcastle2003}, the RHS in Eq.~\eqref{eq:rel_factor} falls in the range between 0.48 ($\theta_{\rm obs}=20^\circ$) and 0.85 ($\theta_{\rm obs}=50^\circ$). 

We should also note that we do not consider the transformation of $w_{\rm ph}$ to the jet comoving frame. This would require information about the angular distribution of the target photon field. For example, if the photon field is isotropic in the laboratory frame then its energy density transforms to the jet co-moving frame as ${\cal T}=\Gamma^2\left(1+\nicefrac{\beta^2}3\right)$, which falls between 1.4 ($\theta_{\rm obs}=50^\circ$) and $2$ ($\theta_{\rm obs}=20^\circ$). If we instead consider the target photons are from a point source at the jet base, then the correction would be ${\cal T} = \Gamma^{-2}(1+\beta)^{-2}$, which is in the range of 2.7 ($\theta_{\rm obs}=50^\circ$) and 4.2 ($\theta_{\rm obs}=20^\circ$). If we take the transformation of $w_{\rm ph}$ into account, the magnetization parameter $\eta_{\B}\di$ would decrease by ${\cal T} $, which strengthen our conclusion that the jet is weakly magnetized. While $\eta_{\rm e,NT}\di$ can be increased by ${\cal T} ^{3/4}$, combined with the correction due to Eq.~(\ref{eq:rel_factor}), the total increase in $\eta_{\rm e,NT}\di$ would be less than a factor of 1.8 for cases considered here.  

\section{Hadronic Scenario}
\label{app:hadron}
Here, we assess the contribution of hadronic processes to the observed VHE emission. We assume that a sub-volume $\cal V$ of the jet produces gamma rays via {\it pp} interactions. The total energy in non-thermal protons in this region is
\begin{equation}
  \begin{split}
    W_{p,\NT} &= w\jet\eta_{p,\NT}{\cal V}\,,\\
    &\simeq 6\times10^{54}\eta_{p,\NT} P_{43}Z_3\beta_{0.5}^{-1}f\vp~{\rm erg}\,,
  \end{split}
\end{equation}
where $f\vp={\cal V}/{V\jet}$ is the filling factor of the production sites. These protons produce VHE gamma-rays on a timescale of $t_{pp} = 10^{15}n^{-1}~{\rm s}$, where $n$ is the gas density in the cgs unit. The luminosity is then
\begin{equation}
  \begin{split}
    L_{\VHE,pp}&\sim\kappa\frac{\chi_{\TeV}W_{p,\NT}}{t_{pp}}\,,\\
    &\sim 10^{39}\eta_{p,\NT}nf\vp P_{43}Z_3\beta_{0.5}^{-1}~\ergs\,,
  \end{split}
\end{equation}
where $\chi_{\TeV}W_{p,\NT}$ is the energy of non-thermal protons in the TeV regime and $\kappa\sim0.17$ is the fraction of the proton energy converted into gamma-rays. To explain the observed luminosity, $7\times10^{38}\,\ergs$, the target density should be very high:

\begin{equation}
    nf\vp \sim 70\left(\frac{\eta_{p, \NT}}{0.1}\right)^{-1} \left(\frac{\chi_{\TeV}}{0.1}\right)^{-1}P_{43}^{-1}Z_3^{-1}\beta_{0.5}\rm \ cm^{-3}.
\end{equation}
If the target gas were involved with the jet motion, the kinetic energy flux would significantly exceed the total jet energy flux (Eq.~\ref{eq:jet_flux}): 
\begin{equation}
\begin{split}
    F_{\rm gas} &= \left(\Gamma -1\right) m_pc^2 nf\vp\beta c \\&\simeq 3\times 10^6f\vp n~{\ergscm},
\end{split}
\end{equation}
Therefore, it is difficult to explain the observed VHE emission by hadronic emission alone. However, some contributions may be possible from the gamma-ray production on dense external cloud or stellar winds \citep{2010ApJ...724.1517B,2012ApJ...755..170B}.

\newpage
\bibliography{cena.bib}
\bibliographystyle{aasjournal}
\vspace{2mm}

\listofchanges
\end{document}